\documentclass[12pt]{article}
\usepackage{amssymb,amsmath,epsfig}
\usepackage{graphicx}
\allowdisplaybreaks

\begin{document}
\title{\bf Stability Analysis of Static Spherical Spacetime in Extended Symmetric Teleparallel Gravity}
\author{M. Zeeshan Gul \thanks{mzeeshangul.math@gmail.com}~,
M. Sharif \thanks {msharif.math@pu.edu.pk}~
and Adeeba Arooj \thanks{aarooj933@gmail.com}\\
Department of Mathematics and Statistics, The University of Lahore,\\
1-KM Defence Road Lahore-54000, Pakistan.}

\date{}
\maketitle

\begin{abstract}
Our manuscript aims to analysis the viability and stability of
anisotropic stellar objects in the modified symmetric teleparallel
gravity. A particular model of this extended theory is considered to
formulate explicit field equations which govern the interaction
between matter and geometry. The configuration of static spherical
symmetric structures is examined through the Finch-Skea solution.
However, the values of unknown constants in the metric potentials
are evaluated by the Darmois junction conditions. For the viability
of proposed stellar objects, the physical parameters including
density, pressure, anisotropy, mass, energy constraints, compactness
function and redshift are analyzed. Furthermore, stability of the
proposed stellar objects is investigated by causality condition,
Herrera cracking approach and adiabatic index. Our findings indicate
that the proposed stellar objects are viable as well as stable in
the presence of correction terms.
\end{abstract}

\section{Introduction}

Einstein's gravitational theory (GR) serves as a cornerstone by
providing a comprehensive description of gravitational field and
matter on cosmic scales. In GR, spacetime is described using
mathematical structures defined by the Riemann's metric. This metric
encodes information about distance and angles in spacetime which
allows physicists to understand the curvature of spacetime caused by
gravity. However, Weyl \cite{4} introduced the concept of a
\emph{length connection} which is different from the standard metric
connection used in Riemannian geometry. Weyl's theory focuses on
gauging the conformal factor adjusting the scale of distances. Weyl
introduced the concept of non-metricity which assures that
divergence of the metric tensor exists. This departure from
Riemannian geometry is a new perspective on the geometric nature of
spacetime. Non-Riemannian geometries incorporate additional
geometric quantities like torsion and non-metricity. Torsion refers
to the twisting or rotation of spacetime, while non-metricity
involves deviations from the standard concept of distance allowing
for a broader understanding of spacetime geometry. Torsion
represents gravitational interaction with teleparallel gravity
instead of GR \cite{4a}. Teleparallel equivalent to GR reformulates
gravity not as a result of spacetime curvature but rather through
torsion. In this framework, the connection used has zero curvature
but non-zero torsion and is still metric-compatible meaning there is
no non-metricity. Symmetric teleparallel gravity provides a
different perspective on gravity, focusing on non-metricity as the
source of gravitational effects in a flat spacetime scenario, i.e.,
zero curvature and zero torsion \cite{4b}. Various extended theories
of gravity in different context has been discussed in
\cite{5a}-\cite{5l}.

Xu et al \cite{4c} introduced $f(\mathrm{Q},\mathrm{T})$  theory by
including the trace of energy-momentum tensor in the action of
symmetric teleparalell theory. This proposal comprises theoretical
implications, compatibility and relevance in cosmology. In recent
research, various studies have delved into the implications of
modified $f(\mathrm{Q},\mathrm{T})$ gravity. Fajardo \cite{21}
emphasized that this modified theory offers alternatives to the
standard cosmological model. Arora et al \cite{22} conducted a study
where they constrained several free parameters in two distinct
$f(\mathrm{Q},\mathrm{T})$ models through various energy conditions.
Their analysis shed light on the viability of this theory, paving
way for a new approach to understand the dark sector of the
universe. Tayde et al \cite{6} explored the feasibility of
traversable wormholes through strange matter in
$f(\mathrm{Q},\mathrm{T})$ background. Pradhan et al \cite{7}
discussed various physical characteristics of gravastars such as
proper length, energy, entropy and surface energy density in this
framework and found a stable gravastar model. Bourakadi et al
\cite{8} conducted a comprehensive study on the black holes in this
gravity. Loo et al \cite{9} investigated $f(\mathrm{Q},\mathrm{T})$
theory with small anisotropy to study the complete cosmic evolution
as our universe is not isotropic since the Planck era. Narawade et
al \cite{10} studied cosmic accelerated expansion in an extended
symmetric teleparallel gravity to understand evolutionary phase of
the universe in modified theory. Khurana et al \cite{13} examined
that this modified theory presents valuable insights into the late
time cosmic acceleration. Shukla et al \cite{14} investigated an
isotropic and homogeneous universe model in this framework by
examining deceleration parameter.

Viable characteristics of stellar objects endorse significant
progress in alternative theories. Ilyas and Ahmad \cite{17} used
observational data from various CS candidates to investigate the
behavior of static spherical structures in the framework of
$f(\mathrm{R})$ gravity. Rej and Bhar \cite{18} employed the
Durgapal-IV metric to analyze the physical characteristics of
anisotropic static spherical solutions in $f(\mathrm{R},\mathrm{T})$
theory. Their examination confirmed that the obtained results remain
in the physically acceptable range. Das \cite{19} delved into stable
spherically symmetric stellar configurations in
$f(\mathrm{R},\mathrm{G})$ gravity. Malik et al \cite{20} analyzed
the charged anisotropic characteristics of CSs using Karmarkar
condition in modified Ricci-inverse gravity. Ditta and Tiecheng
\cite{21a} discussed a comprehensive analysis of the physical
properties of CSs in Rastall teleparallel gravity. Sharif and Gul
\cite{28} checked the stability of proposed CSs using sound speed
method in squared gravity.

Nashed and Capozziello \cite{29} investigated the viable model for
neutron stars in $f(\mathrm{R})$ gravity, suggesting intriguing
possibilities to understand the gravitational behavior of dense
objects. Dey et al \cite{30} explored the viable anisotropic stellar
objects employing the Finch-Skea solutions in
$f(\mathrm{R},\mathrm{T})$ framework. Rej et al \cite{31} conducted
a detailed examination of the charged CSs in the same theoretical
framework, unveiling valuable insights into the interplay between
charge and gravity in compact stellar systems. Kumar et al \cite{32}
investigated crucial insights into the relationship between
$f(\mathrm{R},\mathrm{T})$ gravitational theory and the internal
dynamics of CSs. Maurya et al \cite{33} used the Karmarkar condition
to probe charged relativistic objects in $f(\mathrm{G},\mathrm{T})$
gravity. Shamir and Malik \cite{36} provided valuable insights into
the dynamical stability of compact spherical systems in modified
framework. Lin and Zhai \cite{37} investigated the influence of
effective fluid parameters on the geometry of CSs in the
$f(\mathrm{Q},\mathrm{T})$ theory. The geometry of compact stars
with different considerations in $f(\mathrm{Q})$ and
$f(\mathrm{Q},\mathrm{T})$ theory has been studied in
\cite{038a}-\cite{038g}.

The above literature emphasizes the importance of exploring the
viable attributes of CSs in $f(\mathrm{Q},\mathrm{T})$ theory. This
paper is organized as follows. In section \textbf{2}, we present the
field equations of $f(\mathrm{Q},\mathrm{T})$ gravity and evaluate
the unknown parameters through the Darmois junction conditions.
Section \textbf{3} delves into examining different physical
quantities to identify the physical characteristics exhibited by the
CSs. Section \textbf{4} analyzes the equilibrium state and stability
of the CSs, providing insights into their dynamic behavior. The
findings and implications of our results are given in section
\textbf{5}.

\section{Stellar Objects in $f(\mathrm{Q},\mathrm{T})$ Theory}

The main objective of this section is to provide an outline of the
process by which the field equations are formulated using the
variational principle in modified $f(\mathrm{Q},\mathrm{T})$ theory.
This modified theory has gained significant attention due to its
potential explanations for various cosmological phenomena. The
gravitational action in this framework is altered by the addition of
extra terms that are based on non-traditional geometric measures
apart from the metric tensor. The motivation behind considering this
modified theory lies in the quest for a more comprehensive framework
to describe gravity and cosmic phenomenon.

The action of this theory is expressed as \cite{4c}
\begin{equation}\label{1}
\mathrm{I}=\frac{1}{2}\int \bigg[f(\mathrm{Q},\mathrm{T})+2
\mathrm{L}_{m}\bigg]\sqrt{-g}d^{~4}x,
\end{equation}
where $L_m$ is Lagrangian of matter. The non-metricity and
deformation tensor are defined as
\begin{eqnarray}\label{1a}
\mathrm{Q}&=&-g^{\gamma\eta}(\mathrm{L}^{\lambda}_{~\xi\gamma}
\mathrm{L}^{\xi}_{~\lambda\eta}
-\mathrm{L}^{\lambda}_{~\xi\lambda}\mathrm{L}^{\xi}_{~\gamma\eta}),
\\\nonumber
\mathrm{L}^{\lambda}_{~\xi\gamma}&=&-\frac{1}{2}
g^{\lambda\varsigma} (\nabla_{\gamma}g_{\xi\varsigma}+\nabla_{\xi}
g_{\varsigma\lambda} -\nabla_{\varsigma}g_{\xi\gamma}).
\end{eqnarray}
However, the super-potential of this model is as follow
\begin{equation}\label{1c}
\mathrm{P}^{\lambda}_{~\gamma\eta}=\frac{1}{4}\big[(\mathrm{Q}^{\lambda}
-\bar{\mathrm{Q}}^{\lambda})g_{\gamma\eta}- \delta ^{\lambda}
_{~(\gamma \mathrm{Q}_{\eta})}\big]-\frac{1}{2}\mathrm{L}
^{\lambda}_{~\gamma\eta},
\end{equation}
Although, the relation for $\mathrm{Q}$ is
\begin{equation}\label{1d}
\mathrm{Q}=-\mathrm{Q}_{\lambda\gamma\eta}\mathrm{P}
^{\lambda\gamma\eta}=\frac{1}{4}(\mathrm{Q}^{\lambda\eta\xi}
\mathrm{Q}_{\lambda\eta\xi}-2\mathrm{Q}^{\lambda\eta\xi}
\mathrm{Q}_{\xi\lambda\eta}+2\mathrm{Q}^{\xi}\bar{\mathrm{Q}}_{\xi}-\mathrm{Q}
^{\xi}\mathrm{Q}_{\xi}),
\end{equation}
where
\begin{eqnarray}\nonumber
\mathrm{Q}_{\lambda\gamma\eta}&=&\nabla_{\lambda} g_{\gamma\eta}
=-\partial g_{\gamma\eta,\lambda}+g_{\eta\xi}
\bar{\Gamma}^{\xi}_{~\gamma\lambda}
+g_{\xi\gamma}\bar{\Gamma}^{\xi}_{~\eta\lambda}.
\end{eqnarray}
The calculation of the above relation (\ref{1d}) and its detailed
variation is given in \cite{4c}. Variate the action (\ref{1})
corresponding to $g_{\gamma\eta}$, we have
\begin{eqnarray}\nonumber
\mathrm{T}_{\gamma\eta}&=& \frac{-2}{\sqrt{-g}} \nabla_{\lambda}
(f_{\mathrm{Q}}\sqrt{-g} \mathrm{P}^{\lambda}_{~\gamma\eta})-
\frac{1}{2} f g_{\gamma\eta} + f_{\mathrm{T}}
(\mathrm{T}_{\gamma\eta} + \Theta_{\gamma\eta})
\\\label{1h}
&-&f_{\mathrm{Q}} (\mathrm{P}_{\gamma\lambda\xi}
\mathrm{Q}_{\eta}^{~~\lambda\xi}
-2\mathrm{Q}^{\lambda\xi}_{~~\gamma} \mathrm{P}_{\lambda\xi\eta}),
\end{eqnarray}
where $f_{\mathrm{T}}$ and $f_{\mathrm{Q}}$ are the partial
derivatives corresponding to $\mathrm{T}$ and $\mathrm{Q}$.

Consider a static spherical spacetime as the interior region of
stellar objects to examine the geometry of CSs, defined as
\begin{equation}\label{2}
ds^{2}=dt^{2}e^{\vartheta(r)}-dr^{2}e^\varpi{(r)}-r^{2}d\Omega^{2},
\end{equation}
where $d\Omega^{2}=d\theta^{2}+\sin^{2}\theta d\phi^{2}$. The
anisotropic matter distribution is given by
\begin{equation}\label{3}
\mathrm{T}_{\gamma\lambda}=\mathrm{U}_{\gamma}\mathrm{U}_{\lambda}
\varrho + \mathrm{V}_{\gamma}\mathrm{V}_{\lambda}
p_{r}-p_{t}g_{\gamma\lambda} +
\mathrm{U}_{\gamma}\mathrm{U}_{\lambda}p_{t} -
\mathrm{V}_{\gamma}\mathrm{V}_{\lambda}p_{t},
\end{equation}
where $\mathrm{V}_{\gamma}$ is 4-vector and $\mathrm{U}_{\gamma}$ is
4-velocity of the fluid. The matter-Lagrangian is important in
various cosmic phenomena as it demonstrates the configuration of
fluid in spacetime. The particular value of matter-Lagrangian can
yields significant insights. The well known used formulation of the
matter-Lagrangian in the literature is
$\mathrm{L}_{m}=-\frac{p_{r}+2p_{t}}{3}$\cite{49}. The modified
field equations for static spherical spacetime become
\begin{eqnarray}\nonumber
\varrho&=&\frac{1}{2r^{2}e^{\varpi}}\bigg[2r\mathrm{Q}'f_{\mathrm{Q}
\mathrm{Q}}(e^{\varpi}-1)
+f_{\mathrm{Q}}\big((e^{\varpi}-1)(2+r\vartheta')+(e^{\varpi}+1)r\varpi'
\big)
\\\label{31}
&+&fr^{2}e^{\varpi}\bigg]-\frac{1}{3}f_{\mathrm{T}}(3\varrho+p_{r}+2p_{t}),
\\\nonumber
p_{r}&=&\frac{-1}{2r^{2}e^{\varpi}}\bigg[2r\mathrm{Q}'f_{\mathrm{Q}
\mathrm{Q}}(e^{\varpi}-1)
+f_{\mathrm{Q}}\big((e^{\varpi}-1)(2+r\vartheta'+r\varpi')-2r\vartheta'\big)
\\\label{32}
&+&fr^{2}e^{\varpi}\bigg]+\frac{2}{3}f_{\mathrm{T}}(p_{t}-p_{r}),
\\\nonumber
p_{t}&=&\frac{-1}{4re^{\varpi}}\bigg[-2r\mathrm{Q}'\vartheta'f_{\mathrm{Q}
\mathrm{Q}}
+f_{\mathrm{Q}}\big(2\vartheta'(e^{\varpi}-2)-r\vartheta'^{2}+\varpi'(2e^{\varpi}+r\vartheta')
\\\label{33}
&-&2r\vartheta''\big)+2fre^{\varpi}\bigg]+\frac{1}{3}f_{\mathrm{T}}
(p_{r}-p_{t}).
\end{eqnarray}

The field equations are complicated because of multivariate
functions and their derivatives. We take a specific model as
\cite{50}
\begin{eqnarray}\label{7}
f(\mathrm{Q},\mathrm{T})=\alpha\mathrm{Q}+\beta\mathrm{T},
\end{eqnarray}
where $\alpha$ and $\beta$ are model parameters. This model enhances
our ability to explain gravitational interactions and their
connection with matter and energy. This model provides more accurate
predictions for mysterious components of the universe phenomena
through the refining of mathematical framework and introduction of
new dynamical mechanisms. Furthermore, this  model stands as a
pivotal pursuit in theoretical physics, aiming to unravel the
fundamental essence of physical phenomena from the smallest to the
largest scales. It emerges from the aspiration for a unified
framework capable of elegantly encompassing a wide array of
phenomena spanning cosmology to particle physics. The resulting
field equations becomes
\begin{eqnarray}\nonumber
\varrho&=&\frac{\alpha
e^{-\varpi}}{12r^2(2\alpha^{2}+\beta-1)}\bigg[ \beta(2r(-\varpi'(r
\vartheta'+2)+2r\vartheta''+\vartheta'(r \vartheta'+4))-4e^{\varpi}
\\\label{8}
&+&4)+3\beta r(\vartheta'(4-r \varpi'+r
\vartheta')+2r\vartheta'')+12 (\beta-1)(r
\varpi'+e^{\varpi}-1)\bigg],
\\\nonumber
p_{r}&=&\frac{\alpha e^{-\varpi}}{12r^2(2\beta^{2}+\beta-1)}\bigg[
2\beta\big(r\varpi'(r
\vartheta'+2)+2(e^{\varpi}-1)-r(2r\vartheta''+\vartheta'(r
\vartheta'
\\\nonumber
&+&4))\big)+3\big(r\big(\beta \varpi'(r \vartheta'+4)-2\beta r
\vartheta''-\vartheta'(-4\beta+\beta r \vartheta'+4)\big)-4(\beta-1)
\\\label{9}
&\times&(e^{\varpi}-1)\big)\bigg],
\\\nonumber
p_{t}&=&\frac{\alpha e^{-\varpi}}{12r^2(2\beta^{2}+\beta-1)}\bigg[
2\beta\big(r\varpi'(r
\vartheta'+2)+2(e^{\varpi}-1)-r(2r\vartheta''+\vartheta'
\\\nonumber
&\times&(r \vartheta'+4))\big)+3\big(r \big(2(\beta-1)r
\vartheta''-((\beta-1)r \vartheta'-2)(\varpi'-\vartheta')\big)
\\\label{10}
&+&4\beta(e^{\varpi}-1)\big)\bigg].
\end{eqnarray}

Here, the metric functions ($\vartheta$, $\varpi$) must be finite
and non-singular to obtain the singular free spacetime. In this
regard, we consider Finch Skea solutions which are considered as the
significant tool to find the exact viable solutions for interior
spacetime, defined as \cite{51}
\begin{eqnarray}\label{11}
e^{\vartheta(r)}=(x+\frac{1}{2}yr\sqrt{zr^{2}})^{2},\quad
e^{\varpi(r)}=1+zr^{2},
\end{eqnarray}
The arbitrary constants are denoted by $x$, $y$ and $z$,
respectively. We can evaluate the values of unknown constants by
Darmois junction conditions. Also, we consider spherically symmetric
vacuum solution as the exterior spacetime. The exterior spacetime is
given by
\begin{eqnarray}\label{12}
ds^{2}_{+}=dt^{2}\aleph-d r^{2}\aleph^{-1}-r^{2}d\Omega^{2},
\end{eqnarray}
where $\aleph=\left(1-\frac{2m}{r}\right)$. The continuity of metric
coefficients at the surface boundary $(r=\mathcal{R})$ gives
\begin{eqnarray}\nonumber
g_{tt}&=&(x+\frac{1}{2}y\mathcal{R}\sqrt{z\mathcal{R}^{2}})^{2}=\aleph,
\\\nonumber
g_{rr}&=&1+z\mathcal{R}^{2}=\aleph^{-1},
\\\nonumber
g_{tt,r}&=&y\mathcal{R}\sqrt{z}(x+\frac{1}{2}
y\mathcal{R}\sqrt{z\mathcal{R}^{2}})=\frac{m}{\mathcal{R}^{2}}.
\end{eqnarray}
Solution of these equation gives
\begin{eqnarray}\nonumber
x=\frac{2\mathcal{R}-5m}{2\sqrt{\mathcal{R}^{2}-2m\mathcal{R}}},
\quad y=\frac{1}{\mathcal{R}}\sqrt{\frac{m}{2\mathcal{R}}}, \quad
z=\frac{2m}{\mathcal{R}^{2}(\mathcal{R}-2m)}.
\end{eqnarray}
These constants are important to comprehend the interior of the CSs.
The mass and radius values for the considered stellar objects are
presented in Table \textbf{1}, while the associated constants can be
found in Table \textbf{2}. In the analysis of stellar objects, it is
essential to examine the behavior of metric elements to ascertain
the smoothness and absence of singularities in the spacetime. The
graphical representation depicted in Figure \textbf{1} serves as a
crucial tool in this evaluation process. It is clear that both
metric components display consistent patterns and show an increasing
trend. This behavior is significant as it indicates the absence of
any abrupt or irregular fluctuations in the spacetime metrics
associated with the stellar objects under consideration. Thus, based
on the graphical analysis presented in Figure \textbf{1}, we can
assert that the spacetime appears to be smooth and devoid of
singularities, meeting the required criteria for our investigation.
\begin{table}\caption{Values corresponding to input parameters}
\begin{center}
\begin{tabular}{|c|c|c|}
\hline CSs & $m_{\odot}$ & $\mathcal{R}(km)$
\\
\hline  Her X - 1 \cite{53} & 0.85 $\pm$ 0.15 & 8.1 $\pm$ 0.41
\\
\hline  EXO 1785-248 \cite{54} & 1.30 $\pm$ 0.2 & 10.10 $\pm$ 0.44
\\
\hline  SAX J1808.4-3658 \cite{55} & 0.9 $\pm$ 0.3 & 7.951 $\pm$ 1.0
\\
\hline  4U 1820-30 \cite{56} & 1.58  $\pm$ 0.06 & 9.1 $\pm$ 0.4
\\
\hline  Cen X-3 \cite{58} & 1.49 $\pm$ 0.08 & 9.178 $\pm$ 0.13
\\
\hline  SMC X-4  \cite{53} & 1.29 $\pm$ 0.05 & 8.831 $\pm$ 0.09
\\
\hline  PSR J1903+327 \cite{59} & 1.667 $\pm$ 0.021 & 9.48 $\pm$
0.03
\\
\hline  LMC X-4 \cite{58} & 1.04 $\pm$ 0.09 & 8.301 $\pm$ 0.2
\\
\hline
\end{tabular}
\end{center}
\end{table}
\begin{table}\caption{Values of unknown parameters.}
\begin{center}
\begin{tabular}{|c|c|c|c|}
\hline CSs & $x$ & $y$ & $z$
\\
\hline  Her X - 1 & -0.924111 & 0.0343333 & 0.00682713
\\
\hline EXO 1785-248 & -0.908173 & 0.0304946 & 0.00599414
\\
\hline  SAX J1808.4-3658 & -0.918476 & 0.0363264 & 0.00792193
\\
\hline  4U 1820-30 & -0.881827 & 0.0393097 & 0.0126622
\\
\hline  Cen X-3 & -0.887784 & 0.0376881 & 0.0108966
\\
\hline  SMC X-4 & -0.89724 & 0.0371547 & 0.00969829
\\
\hline  PSR J1903+327 & -0.880727 & 0.0379742 & 0.0119768
\\
\hline \hline  LMC X-4 & -0.910411 & 0.0366062 & 0.00849917
\\
\hline
\end{tabular}
\end{center}
\end{table}
\begin{figure}
\epsfig{file=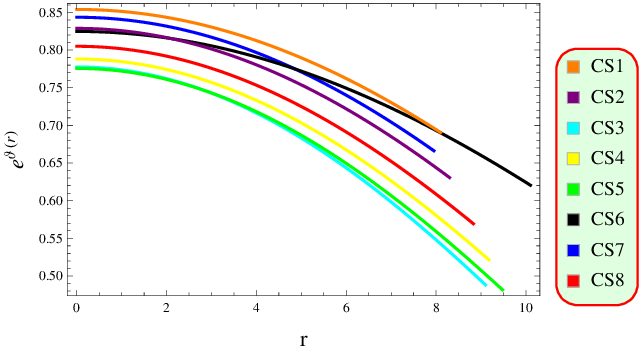,width=.5\linewidth}
\epsfig{file=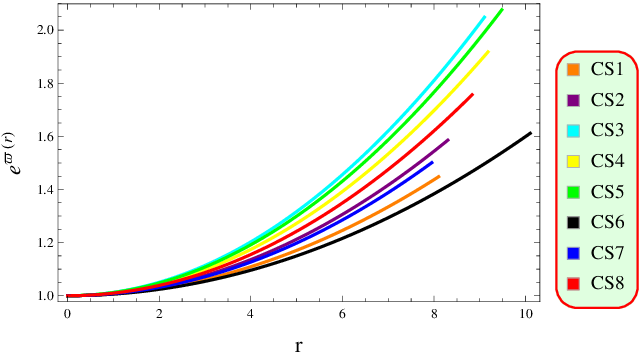,width=.5\linewidth}\caption{Plot of metric
potentials.}
\end{figure}

\section{Viable Characteristics of Stellar Objects}

In this section, we examine the viable features of CSs using
graphical analysis. For viable and stable CSs with a specific
radius, the given conditions are required to be satisfied.
\begin{itemize}
\item
The metric coefficients need to be monotonically increasing and
non-singular at the center, which ensures that spacetimes does not
contain any kind of irregularities.
\item
The behavior of matter contents should be monotonically decreasing
and $p_{r}(r=\mathcal{R})=0$ to assure that the CSs has a stable
denser core.
\item
The matter gradient must vanish at the core and then demonstrate
negative behavior towards the boundary.
\item
Positive energy bounds ensure the presence of normal matter in the
stellar objects, which is necessary for viable geometry of CSs.
\item
The EoS parameters must fall in the range of [0,1] for stellar
structures to be viable.
\item
The mass function must be continuous at the core and then shows
positively increasing behavior.
\item
The compactness and redshift functions must be less than
$\frac{4}{9}$ and 5.21, respectively, for viable geometry of CSs.
\item
The forces must satisfy equilibrium condition to maintain the
stability.
\item
For CSs to be stable, the velocities of sound speed should remain in
the range of [0,1], while the adiabatic index must be greater than
1.33.
\end{itemize}
These constraints provide a framework to understand the behavior of
CSs and ensure that their properties are consistent.

\subsection{Graphical Analysis of Fluid Parameters}

The investigation of fluid parameters such as density and pressure
is essential to understand the internal features of neutron stars.
These matter variables are anticipated to be maximum at the core due
to their intense density, counteracting gravitational forces and
maintaining the stability of CSs against collapse. The corresponding
field equations are given as follows
\begin{eqnarray}\nonumber
\varrho&=&\alpha\bigg[-3yzr^{3}(3+zr^{2})+2yr(15+zr^{2}(13+zr^{2}))\beta+2x(3+zr^{2})\sqrt{zr^{2}}
\\\label{13}
&\times&(2\beta-3)\bigg]\bigg[3(1+zr^{2})^2(yzr^{3}+2x\sqrt{zr^{2}})(\beta-1+2\beta^{2})\bigg]^{-1},
\\\nonumber
p_{r}&=&\alpha z
\bigg[2x\sqrt{zr^{2}}(3+zr^{2}(2\beta-3)+6\beta)-yr(6(2+\beta)+zr^{2}(9-10\beta+z
\\\label{14}
&\times&r^{2}(2\beta-3)))\bigg]\bigg[3(1+zr^{2})^2(yzr^{3}+2x\sqrt{zr^{2}})(\beta-1+2\beta^{2})\bigg]^{-1},
\\\nonumber
p_{t}&=&\alpha z
\bigg[2x\sqrt{zr^{2}}(3+(6+4zr^{2})\beta)+yr(-6(2+\beta)+zr^{2}(-3+(4zr^{2}-2)
\\\label{15}
&\times&\beta))\bigg]\bigg[3(1+zr^{2})^2(yzr^{3}+2x\sqrt{zr^{2}})(\beta-1+2\beta^{2})\bigg]^{-1}.
\end{eqnarray}
\begin{figure}
\epsfig{file=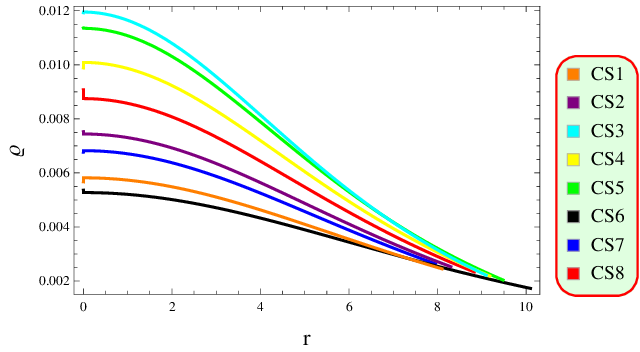,width=.5\linewidth}
\epsfig{file=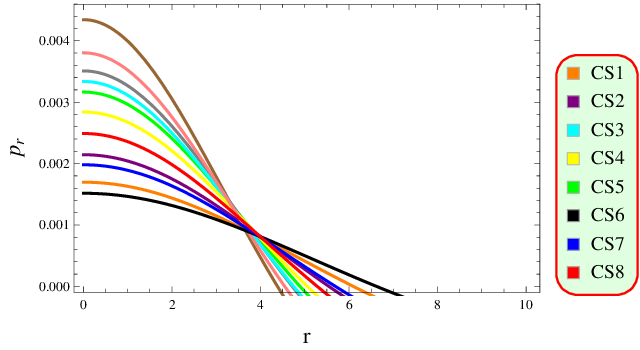,width=.5\linewidth}\center
\epsfig{file=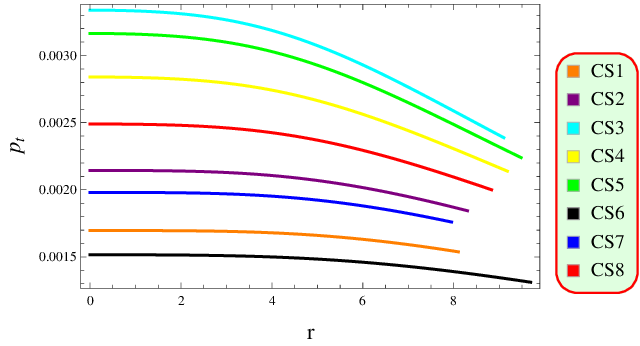,width=.5\linewidth}\caption{Plot of fluid
variables.}
\end{figure}
\begin{figure}
\epsfig{file=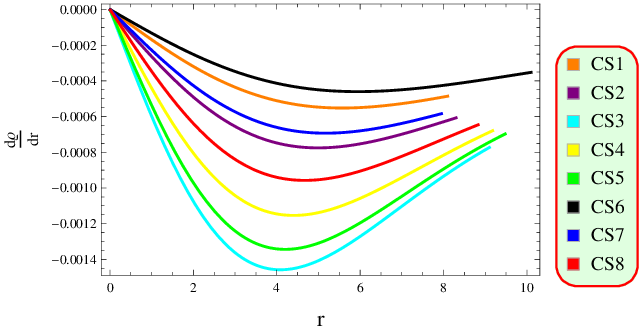,width=.5\linewidth}
\epsfig{file=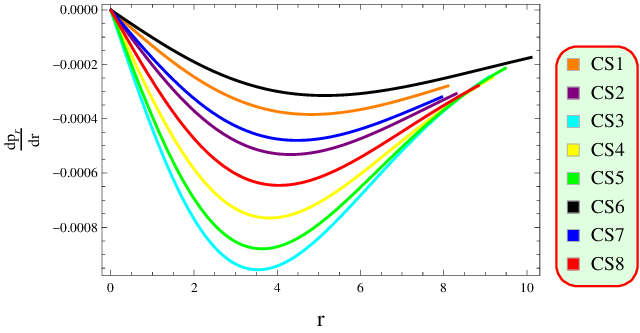,width=.5\linewidth}\center
\epsfig{file=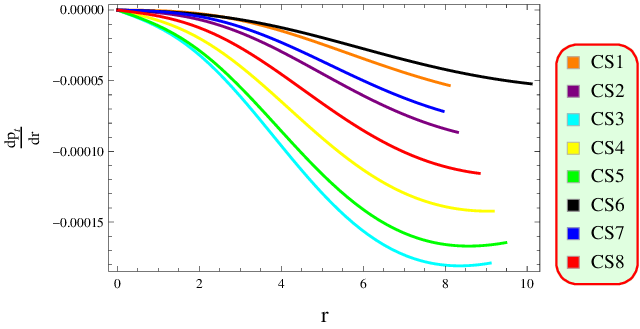,width=.5\linewidth}\caption{Plot  of gradient of
fluid variables.}
\end{figure}
The plots in Figure \textbf{2} determine that the matter contents
are maximum at the core before decreasing, highlighting the dense
nature of the CSs. Additionally, the radial pressure in the
considered CSs shows a consistent decrease as distance from the
center increases until it dissipates at the boundary. Figure
\textbf{3} manifests that the CSs have highly dense structures in
this framework as the gradient of density and pressure components
vanish at the core and become negative thereafter.

\subsection{Behavior of Anisotropy}

An anisotropic fluid refers to a fluid that exhibits different
physical properties or behavior in different directions. The term
``anisotropic'' comes from the Greek words ``aniso'' meaning unequal
or different and ``tropos'' meaning direction. Anisotropy refers to
a difference in pressure along different directions in the system.
The pressure in a star is isotropic when there are no additional
forces or anisotropic effects present. However, in certain
situations such as the presence of strong magnetic fields or other
factors, pressure becomes anisotropic. One example of anisotropy is
the gravitational field around a rotating object. The gravitational
field surrounded by rotating massive objects such as a spinning
black hole is not uniform in all directions. The gravitational
attraction is stronger in some directions than in others, resulting
in anisotropic effects. This phenomenon is known as frame-dragging,
where the rotation of the object drags the surrounding spacetime
along with it. Figure \textbf{4} indicates the existence of a
repulsive force as the behavior of anisotropy is positive, which
plays a crucial role in sustaining large-scale structures and
preventing gravitational collapse.
\begin{figure}\center
\epsfig{file=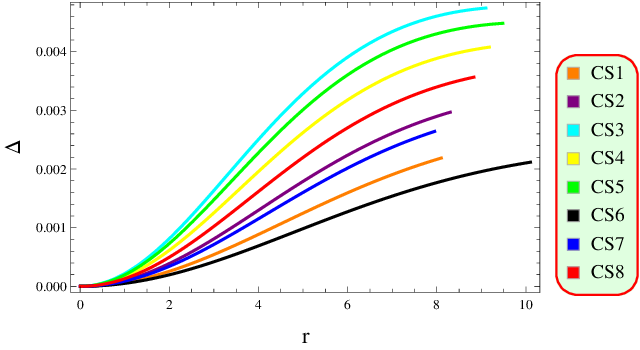,width=.5\linewidth}\caption{Plot of Anisotropy.}
\end{figure}

\subsection{Energy Constraints}

Astrophysical entities are composed of a variety of materials in
their composition and it is important to differentiate the types of
matter (exotic/ordinary) present in the celestial objects. Energy
constraints are necessary to examine the viable fluid configurations
in the system. These limitations play a crucial role in
investigating the presence of specific cosmic formations and
understanding how matter and energy interact under the influence of
gravity. These constraints manifest the physical viability of the
matter configuration in the neutron stars. The energy conditions are
characterized into four types as
\begin{itemize}
\item Null Energy Condition\\\\
$0\leq \varrho+p_{r}$, \quad $0\leq \varrho+p_{t}$.
\item Dominant Energy Condition\\\\
$0\leq \varrho\pm p_{r}\geq 0$, \quad  $0\leq \varrho\pm p_{t}$.
\item Weak Energy Condition\\\\
$0\leq \varrho+p_{r}+\geq 0$,\quad $0\leq \varrho+p_{t}$, \quad
$0\leq \varrho$.
\item Strong Energy Condition\\\\
$0\leq \varrho+p_{r}$, \quad $0\leq \varrho+p_{t}$, \quad $0\leq
\varrho+p_{r}+2p_{t}$.
\end{itemize}
Scientists can gain insights into the nature and behavior of cosmic
structures by analyzing these energy bounds and their effects on the
stress-energy tensor, contributing to our understanding of the
dynamics and evolution of the universe. Figure \textbf{5} shows that
the considered CSs are viable as all energy constraints are
satisfied in the presence of modified terms.
\begin{figure}
\epsfig{file=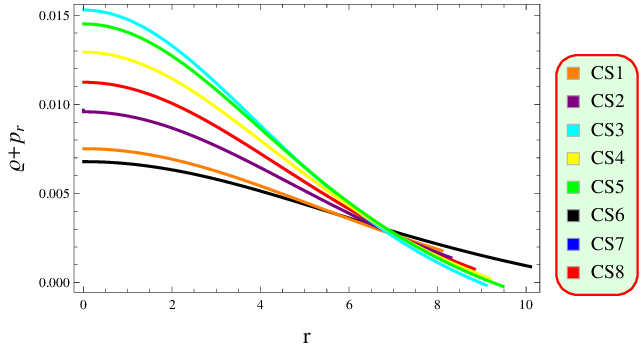,width=.5\linewidth}
\epsfig{file=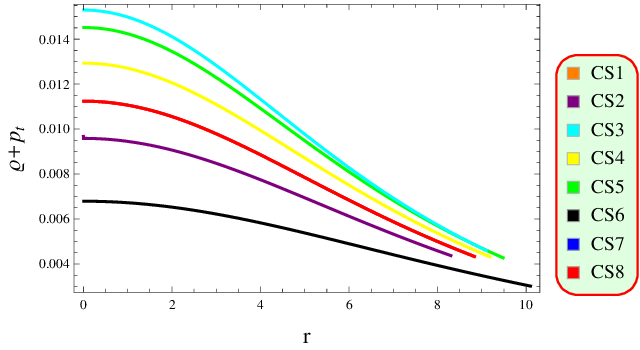,width=.5\linewidth}
\epsfig{file=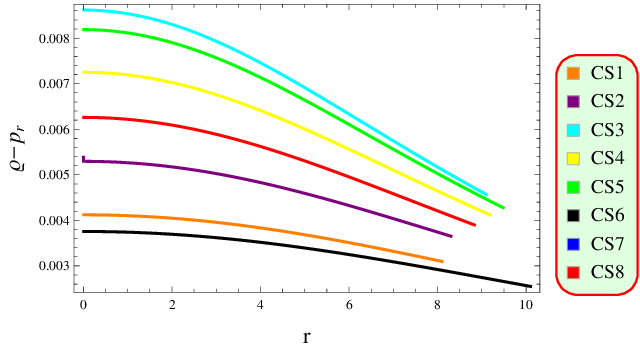,width=.5\linewidth}
\epsfig{file=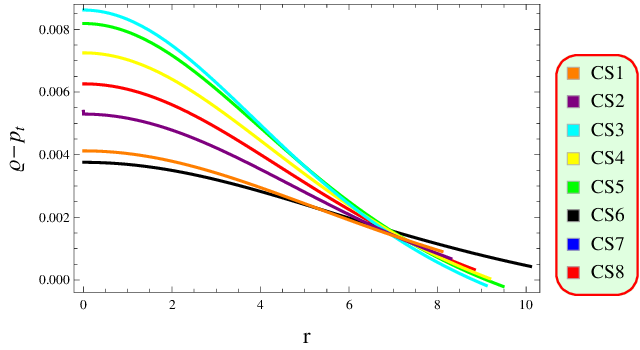,width=.5\linewidth}\caption{Plot of energy
constrains.}
\end{figure}

\subsection{Evolution of State Parameters}

The EoS parameters explain how energy density is related to
anisotropic pressure in different types of systems. The radial
component $(\omega_{r}=\frac{p_{r}}{\varrho})$ and transverse
component $(\omega_{t}=\frac{p_{t}}{\varrho})$ of EoS parameters
must satisfy range [0,1] for viable stellar stars \cite{60}. Using
Eqs.(\ref{13})-(\ref{15}), we have
\begin{eqnarray}\nonumber
\omega_{r}&=&\bigg[2x\sqrt{zr^{2}}(3+zr^{2}(3-2\beta)+6\beta)-yr(6(2+\beta)+zr^{2}(9-10\beta
\\\nonumber
&+&zr^{2}(2\beta-3)))\bigg]\bigg[2x\sqrt{zr^{2}}(3+zr^2)(2\beta-3)+y(30r\beta+z^{2}r^{5}
\\\nonumber
&\times&(2\beta-3)+zr^{3}(26\beta-9))\bigg]^{-1},
\\\nonumber
\omega_{t}&=&\bigg[2x\sqrt{zr^{2}}(3+(6+4zr^{2})\beta)+yr(-6(2+\beta)+zr^{2}(-3+(4zr^{2}
\\\nonumber
&-&2)\beta))\bigg]\bigg[2x\sqrt{zr^{2}}(3+zr^2)(2\beta-3)+y(30r\beta+z^{2}r^{5}(2\beta-3)
\\\nonumber
&+&zr^{3}(26 \beta-9))\bigg]^{-1}.
\end{eqnarray}
Figure \textbf{6} manifests that the behavior of EoS parameters
satisfy the viability condition corresponding to all considered
stellar objects.
\begin{figure}
\epsfig{file=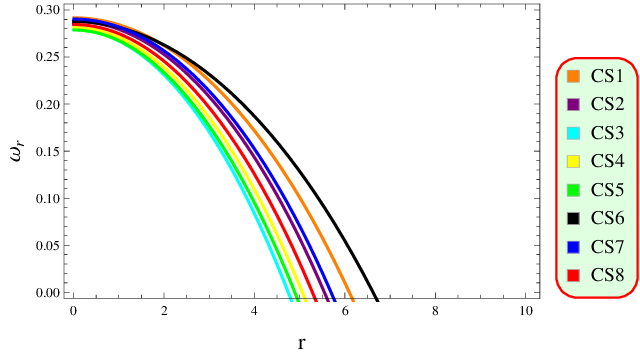,width=.5\linewidth}
\epsfig{file=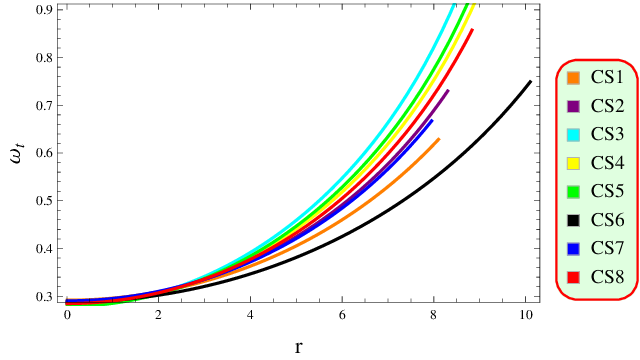,width=.5\linewidth}\caption{Pots of EoS
parameters.}
\end{figure}

\subsection{\textbf{Evolution of Different Physical Properties}}

\begin{figure}\center
\epsfig{file=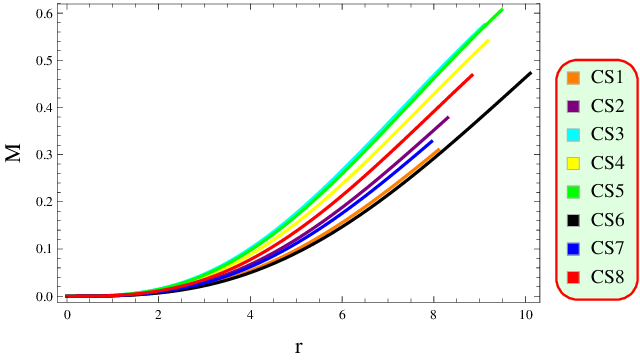,width=.5\linewidth} \caption{Behavior of mass
function.}
\end{figure}
The mass of stellar objects is described as
\begin{equation}\nonumber
M=4\pi\int^{\mathcal{R}}_{0} r^{2}\varrho dr.
\end{equation}
Figure \textbf{7} demonstrates that the mass function is
monotonically increasing and $M\rightarrow 0$ as $r\rightarrow 0$,
suggesting that there are no irregularities in the mass
distribution. Various physical characteristics can be assessed to
analyze the structural configuration of cosmic objects. A
fundamental factor in assessing the viability of CSs is the
compactness function, represented by $(u=\frac{M}{r})$. This
function offers insights into the distribution of mass relative to
the radius of a CS and its concentration. The compactness factor is
a physical parameter that provides a quantitative measure of how
densely packed mass is within a given radius. There is a specific
limit for the compactness function proposed by Buchdhal for a
physically relevant model \cite{61}. According to his criterion, the
mass-radius ratio should be less than $4/9$ for viable stellar
objects.

The surface redshift is a significant factor as it provides
important information about the brightness and energy of light
emitted from their surfaces, which is caused by the gravitational
redshift due to the strong gravity. It is a phenomenon which
explains the change in frequency (wavelength) of light or other
electromagnetic radiations as it travels away from a gravitational
field. As the light moves away from the gravitational field, it
loses energy and thus its wavelength is increased, causing it to
shift towards the red end of the electromagnetic spectrum. It is
denoted by $Z_{s}$ and mathematically expressed as
\begin{equation}\label{31}
Z_s =- 1+\frac{1}{\sqrt{1-2u}}.
\end{equation}
In case of anisotropic configuration, the redshift at the surface
must satisfy the specific condition as $Z_{s}<5.211$ for CSs to be
viable \cite{62}. The graphs in Figure \textbf{8} demonstrate that
both compactness and redshift functions meet the essential
feasibility criteria.
\begin{figure}
\epsfig{file=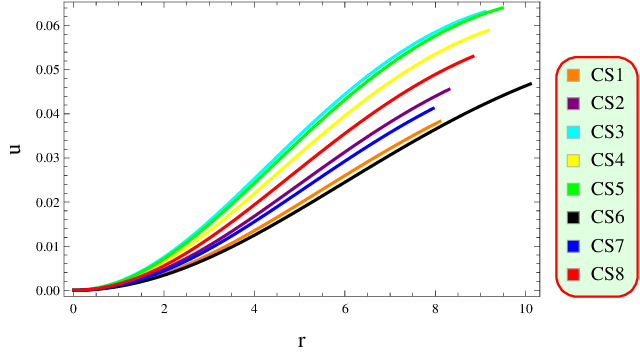,width=.5\linewidth}
\epsfig{file=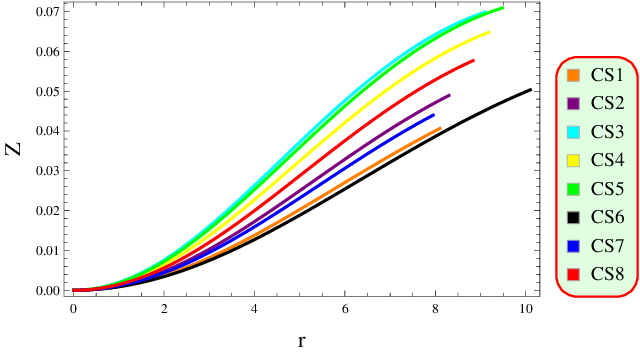,width=.5\linewidth}\caption{Plot of compactness
and redshift functions.}
\end{figure}

\section{Stability Analysis}

It is important to comprehend the behavior and physical
characteristics of celestial objects in the field of gravitational
physics. The stability of cosmic formation is significant to develop
their reliability and coherence. Scientists have investigated the
conditions that determine the stability of these formations against
various forms of oscillations. To assess the stability of pulsars,
researchers use the methods such as causality constraint and Herrera
cracking approach and adiabatic index provide important perspectives
on the structural integrity of astronomical objects.

\subsection{Causality Condition}

The stability of CSs can be evaluated by considering the causality
constraint, which states that nothing can travel faster than the
speed of light. In order to maintain stable configurations, both the
radial and tangential velocities of sound ($v_r =
\frac{dp_r}{d\varrho}$ and $v_t = \frac{dp_t}{d\varrho}$) must fall
in the range of 0 to 1 \cite{63}. These characteristics related to
sound speed play a critical role in ensuring the stability of CSs.
These are given as follows
\begin{eqnarray}\nonumber
v_{r}&=&\bigg[4x^{2}zr(3+zr^{2}(3-2\beta)+14\beta)-4xy\sqrt{zr^{2}}(6+8\beta+zr^{2}(-16\beta
\\\nonumber
&+&3+zr^{2}(2\beta-3)))-y^{2}r(6(2+\beta)+zr^{2}(18(2+\beta)+zr^{2}(-22\beta
\\\nonumber
&+&21+zr^{2}(2\beta-3))))\bigg]\bigg[4x^{2}zr(5+zr^{2})(2\beta-3)+4xy\sqrt{zr^{2}}(20\beta
\\\nonumber
&+&z^{2}r^{4}(2\beta-3)+5cr^{2}(-3+4\beta))+y^{2}r(30\beta+zr^{2}(90\beta+zr^{2}(-15
\\\nonumber
&+&50\beta+zr^{2}(2\beta-3))))\bigg]^{-1},
\\\nonumber
v_{t}&=&2\bigg[4x^{2}zr(3+2(2+zr^{2})\beta)+2xy\sqrt{zr^{2}}(-9-2\beta+zr^{2}(3+4
\\\nonumber
&\times&(1+zr^{2})\beta))+y^{2}r(-3(2+\beta)+zr^{2}(-9(2+\beta)+zr^{2}(-3+2
\\\nonumber
&\times&(-2+zr^{2})\beta)))\bigg]\bigg[4x^{2}zr(5+zr^{2})(2\beta-3)+4xy\sqrt{zr^{2}}(20\beta+z^{2}
\\\nonumber
&\times&r^{4}(2\beta-3)+5cr^{2}(-3+4\beta))+y^{2}r(30\beta+zr^{2}(90\beta+zr^{2}(-15
\\\nonumber
&+&50\beta+zr^{2}(2\beta-3))))\bigg]^{-1}.
\end{eqnarray}
Figure \textbf{9} shows that the considered CSs satisfy the required
condition. Thus, this modified theory supports the existence of
physically viable and stable CSs.
\begin{figure}
\epsfig{file=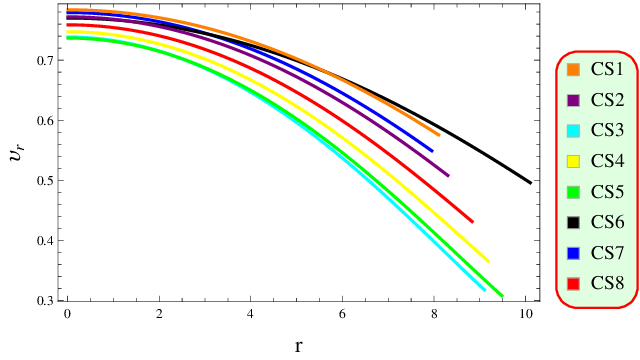,width=.5\linewidth}
\epsfig{file=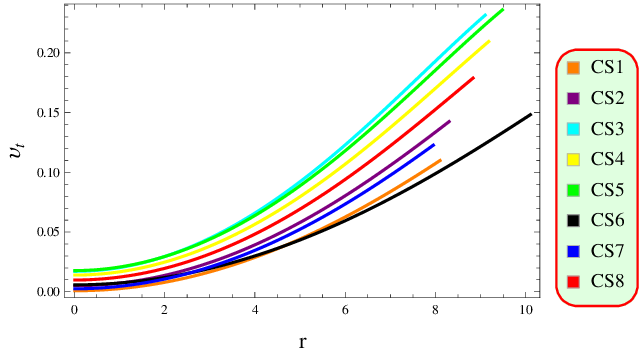,width=.5\linewidth}\caption{Plot of causality
condition.}
\end{figure}

\subsection{Herrera Cracking Approach}

The analysis of solution's stability is based on a mathematical
method called the cracking approach $(0\leq\mid
v_{t}-v_{r}\mid\leq1)$, which was developed by Herrera \cite{64}.
Satisfying this condition indicates stable cosmic structures capable
of long-term existence, otherwise, it signifies instability and will
collapse. This method enables researchers to determine the stability
of cosmic structures, which is essential for understanding their
behavior in the universe. Figure \textbf{10} depicts the fulfillment
of the cracking condition as the both radial and tangential sound
speed components lie in the range [0,1], which ensures the stability
of the stellar objects under consideration.
\begin{figure}\center
\epsfig{file=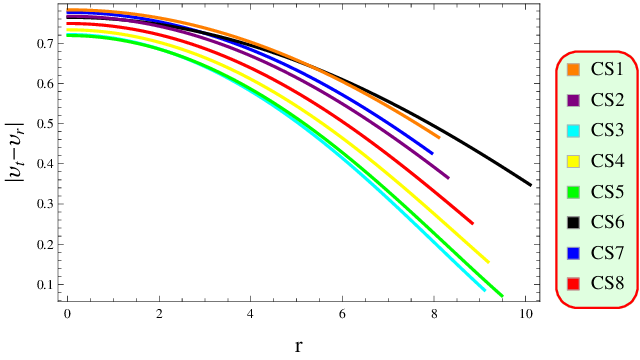,width=.5\linewidth}\caption{Plot of Herrera
cracking approach.}
\end{figure}

\subsection{Adiabatic Index}

This method is considered as a significant method to determine the
stability of cosmic objects, providing insights into their
composition and behavior. This characterizes how pressure changes
corresponding to density variations in stars which plays a pivotal
role in astrophysics, described as
\begin{eqnarray}\nonumber
\Gamma_{r}=\frac{\varrho+p_{r}}{p_{r}}v_{r},\quad
\Gamma_{t}=\frac{\varrho+p_{t}}{p_{t}}v_{t}.
\end{eqnarray}
It is essential to determine the value of $\Gamma$ for the stability
analysis. A stable object exhibits that the values of $\Gamma$
should be greater than 4/3, while instability leading to collapse
occurs when the value falls below this limit. Figure \textbf{11}
shows that our system remains stable in the presence of correction
terms.
\begin{figure}
\epsfig{file=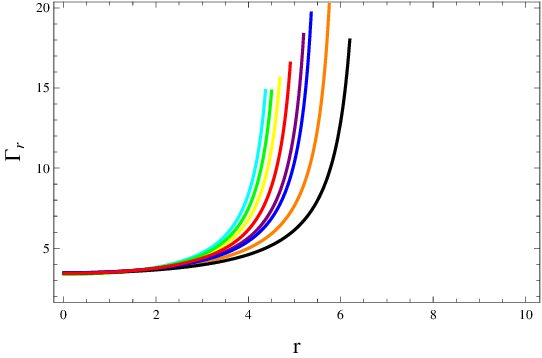,width=.5\linewidth}
\epsfig{file=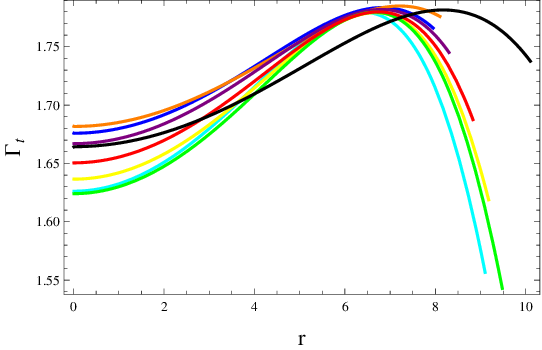,width=.5\linewidth}\caption{Plot of adiabatic
index.}
\end{figure}

\section{Conclusions}

This research investigates the feasibility and stability of CSs in
the extended symmetric teleparallel theory. The main aim is to
investigate whether incorporating non-metricity into the
gravitational field equations results in feasible solutions for CSs.
Introducing these terms to the theory presents new insights into how
matter and geometry interact under extreme gravitational
circumstances. Furthermore, we have conducted a graphical analysis
of various physical characteristics to verify the viability of the
CSs in the proposed theoretical framework. In addition, we have
assessed stability through methods that incorporate sound speed and
adiabatic index.

The metric functions are essential and have a significant impact in
describing the geometry of spacetime. We have found that metric
coefficients are consistent and nonsingular which ensure that the
spacetime is smooth and devoid of any singularities (Figure
\textbf{1}). This smoothness is a fundamental requirement for any
viable cosmological model. The material contents are higher, regular
and the most concentrated at the center of the observed CSs (Figure
\textbf{2}) which indicate a stable core and this behavior of fluid
parameters is desirable to maintain the structural integrity of the
CSs. Furthermore, the decrease in matter content towards the
boundary suggests a viable distribution in the CSs. The vanishing
radial pressure at the surface boundary ensures the physical
viability of the CSs. The matter contents exhibit a negative
gradient, indicating a dense profile of the suggested stellar
objects (Figure \textbf{3}).

The anisotropy vanishes at the center of CSs which is a desirable
feature for maintaining the stability of the CSs (Figure
\textbf{4}). Moreover, anisotropic pressure is directed outward
which is a crucial feature for compact stellar configurations. All
energy constraints are positive (Figure \textbf{5}), which confirm
the existence of ordinary matter in the interior of CSs. Figure
\textbf{6} indicates the viability of the considered model as the
EoS parameter lies in the range of 0 and 1. The mass function
(Figure \textbf{7}) remains consistent at the center of the CSs and
demonstrates a steady increase as the radial coordinate increases.
This behavior is indicative of a viable mass distribution in the
CSs. The compactness factor being less than 4/9 and the redshift
being less than 5.2 (Figure \textbf{8}) support the viability of the
compact structures. The stability limits guarantee the existence of
stable CSs in this framework (Figures \textbf{9}-\textbf{11}).

It is worthwhile to mention here that the range of physical
quantities in $f(\mathrm{Q},\mathrm{T})$ increases and provides more
viable as well as stable compact stars as compared to GR
\cite{65}-\cite{67} and other modified theories \cite{68}. For
example in $f(\mathrm{R})$ theory, it is found that physical
quantities such as effective matter variables, energy conditions,
EoS parameters and speed of sound are satisfied for very small range
and the compact star Her X-1 under the effects of the second gravity
model does not remain stable \cite{69}. In
$f(\mathrm{G},\mathrm{T})$ theory, it is analyzed that the evolution
of compact stars (SAXJ1808.4-3658 and 4U1820-30) is supported by all
three gravity models while for HerX - 1, there are some restrictions
asserted by the second model to be completely physically viable
\cite{69a}. Additionally, in the framework of
$f(\mathrm{R},\mathrm{T}^{2})$ theory, it is evident that CSs are
neither theoretical viable nor stable at the center
\cite{70}-\cite{70b}. In the light of these findings, we conclude
that all the CSs considered in this study exhibit both physical
viability and stability at their centers in this modified theory.
Consequently, our findings suggest that more viable and stable CSs
can exist in this modified framework.

A new class of interior solutions to the Einstein field equations
for an anisotropic matter distribution using a linear EoS \cite{72}.
They verified the physical acceptability of the solutions by the
current estimated data of CS 4U-160852. In contrast, our study
diverges by adopting the $f(\mathrm{Q},\mathrm{T})$ theory, a
different theoretical framework that allows for a comprehensive
exploration of the effects of modified terms on the viability and
stability of CSs. Using distinct methodologies and techniques, we
have analyzed several anisotropic CSs, discerning their viability
and stability under the influence of modified gravitational
dynamics. Our findings reveal the feasibility of the proposed CSs
even in the presence of modified terms, underscoring the robustness
of our theoretical framework. By expanding the scope of our
investigation to encompass a broader spectrum of CSs, we contribute
to the ongoing discourse on the behavior of CSs in alternative
gravitational theory, thus enriching our understanding of
astrophysical phenomena across diverse theoretical landscapes.
\\\\
\textbf{Data Availability Statement:} No data was used for the
research described in this paper.

\end{document}